# Interface Dynamics at a Four-fluid Interface during Droplet Impact on a Two-Fluid System

*Akash Chowdhury, Sirshendu Misra, Sushanta K. Mitra*[*]

Micro & Nano-Scale Transport Laboratory, Waterloo Institute for Nanotechnology, Department of Mechanical and Mechatronics Engineering, University of Waterloo, 200 University Avenue West, Waterloo, Ontario N2L 3G1, Canada

Email: skmitra@uwaterloo.ca

## Abstract

We investigate the interfacial dynamics involved in the impact of a droplet on a liquid-liquid system, which involves the impingement of an immiscible core liquid drop from a vertical separation onto an interfacial shell liquid layer floating on a host liquid bath. The dynamics have been studied for a wide range of impact Weber numbers and two different interfacial shell liquids of varying volumes. The core drop, upon impact, dragged the interfacial liquid into the host liquid, forming an interfacial liquid column with an air cavity inside the host liquid bath. The dynamics is resolved into cavity expansion and rapid contraction, followed by thinning of the interfacial liquid. The interplay of viscous dissipation, interfacial pull, and core drop inertia influenced the necking dynamics. The viscous dissipation increases with the thickness of the interfacial layer, which depends on its volume and lateral spread over the water. The necking dynamics transitioned from inertia-dominated deep seal closure at higher spread, lower interfacial film volumes, and higher Weber numbers, into inertia-capillary dominated deep seal closure with an increase in film volumes, decrease in the spread of the interfacial fluid or decrease in Weber number, and finally transitioned into a no seal closure at high volumes, low spread, and low Weber numbers.

## Introduction

Free surface flow due to interface deformation during a droplet impact is a common natural phenomenon, such as falling raindrops on ocean or small puddles leading to soil erosion.[1,2] It also has varied industrial applications such as spray coating,[3] spray cooling,[4] inkjet printing,[5] and pesticide spraying.[6] Hence, the impingement of droplets has been extensively investigated on various surfaces, including solid (flat[7]/structured[8]) rigid surfaces,[9–12] soft substrates, [13] thin films supported by a solid surface,[14,15] and liquid pools[16–21] ever since the pioneering work by Worthington.[22] The impact of a liquid drop on a pool of the same liquid leads to the formation of a crater[2] below the pool liquid surface and its subsequent retraction, leading to different interfacial phenomena such as bubble entrainment and Worthington Jet, which have been studied widely.[23–27] The crater dynamics are significantly altered when an immiscible drop is impacted into a liquid pool. The droplet transforms into a thin film spreading along the crater surface, leading to interesting interfacial dynamics along the two-liquid interface.[28–31]

The interface dynamics become further complex if we modify the pool into a two-layered system of immiscible fluids,[32–34] a common occurrence in nature, including raindrops impacting an oil spill-covered seawater. However, two-layer system of immiscible fluid has little reported literature.[32–34,36] A recent study on the impact of an oil drop on a pool of water covered by an oil film investigated the critical thickness of the oil film.[33] Below a critical thickness, both the oil-air surface and the oil-water interface would form a crater, while above the critical thickness, the oil film acts as a pool with no influence of water. In another work, the splash formation on the impact of rainwater drops on seawater covered with oil spills of varied thicknesses was studied.[35] The formation of various types of composite oil-water droplets in various Weber number regimes in a similar scenario was investigated by Wang et al.[32] However, these works use droplet liquid to be the same as the top floating liquid layer[33] or the host liquid.[32] In this work, we study the interfacial dynamics during the impact of a core droplet onto a two-layered system comprising a thin liquid film (thickness ~ O(100-1000 μm)) floating on an underlying liquid pool wherein the impinging drop, the interfacial layer (top floating liquid), and the underlying liquid pool are three different immiscible liquids, making it a four-phase system (including the surrounding air). In our study, the volume of the interfacial layer is chosen to form a thin floating lens instead of a thick liquid column. The system offers us a paradigm to investigate the interface dynamics at a complex four-fluid interface defined by the four participating deformable fluidic entities.

An interesting practical example of droplet impact on the aforementioned two-fluid system could be found in liquid-liquid encapsulation, where stable, ultrafast wrapping of different liquid droplets ("core") is achieved by impacting them on a thin interfacial layer of a compatible shell-forming liquid, stably floating on a host bath.[37–41] In our previous work on liquid-liquid encapsulation, we have demonstrated impact-driven stable wrapping of various core droplets (laser oil[37]/honey[38] /maple syrup[38] /ferrofluids[39,40]) using multiple different shell-forming liquids, (canola oil[37] /PDMS[38]) as an interfacial layer floating on a DI water (host) bath inside a glass cuvette. The core droplet, in the presence of sufficient kinetic energy, penetrated the interfacial layer and settled at the bottom of the glass cuvette while, in the absence of sufficient kinetic energy, it remained trapped in the interfacial layer. In both cases, it was observed to be completely wrapped with the interfacial liquid. These works propose a promising pathway for stable and ultrafast wrapping of multiple target analytes even in aggressive surroundings. However, a complete hydrodynamic understanding of the impact phenomena that precede the formation of the encapsulated droplet is critical to identifying the process parameter for desired core-shell morphology and the stability of encapsulated cargo which is important for the commercialization of the process in pharmaceutical and nutraceutical applications.

In this study, we investigate the impact hydrodynamics of a core droplet on the aforementioned two-fluid system comprising a floating interfacial layer on top of a host bath. Upon impact, the core drop with sufficient kinetic energy is observed to drag a portion of the interfacial liquid into the host bath forming a transient cylindrical cavity that, under hydrostatic pressure and interfacial pull, decreases in diameter like a necking process.[42–48] Following the closure of the cavity, a thin column of interfacial liquid remains, which undergoes a slow thinning process. In this study, the necking dynamic was studied in detail by varying the experimental parameters such as the vertical separation between the drop and interfacial layer(impact height $H$), the interfacial liquid volume ($V_{\text{film}}$), and the thermophysical properties of the interfacial liquid. Time-resolved high-speed micrographs were captured to investigate the necking dynamics.

**Materials and Methods**

The experiment involves three immiscible liquids, defined as the core drop (liquid L1), the interfacial layer (liquid L2), and the host liquid (liquid L3). The interfacial layer forms a floating lens on the host liquid and the core liquid drop is dispensed from an impact height $H$ above the interfacial layer, as shown in Figure 1(a) The resulting dynamics are captured using a high-speed camera. The host liquid is chosen to be deionized (DI) water (purified by Milli-Q, Millipore Sigma, Ontario, Canada) with density $\rho_3 = 1000$ kg/m$^3$, dynamic viscosity $\mu_3 = 1$ mPa $\cdot$ s, and liquid–air surface tension $\gamma_3 = 72$ mN/m. We use two different liquids for an interfacial layer that wraps the core droplet. They are underline{silicone oil} (Product number: 378356, Sigma Aldrich) with density $\rho_2 = 963$ kg/m$^3$, dynamic viscosity $\mu_2 = 48.1$ mPa $\cdot$ s, liquid-air surface tension $\gamma_2 = 20$ mN/m, and liquid-water surface tension $\gamma_{23} = 35.7$ mN/m and underline{mineral oil} (Product number: M8410, Sigma Aldrich) with density $\rho_2 = 877$ kg/m$^3$, dynamic viscosity $\mu_2 = 28.7$ mPa $\cdot$ s, liquid-air surface tension $\gamma_2 = 35$ mN/m and liquid-water surface tension $\gamma_{23} = 25$ mN/m. The liquid-water surface tension $\gamma_{23}$ was measured using pendant drop tensiometry. The core droplet is a particular class of laser liquid – a mixture of silicanes and polyphenol ethers, with a water solubility of 0.1% (Product Code: 57B63, Cargille Laboratories Inc., Cedar Grove, NJ, USA). The relevant material properties of laser oil are as follows: density $\rho_1 = 1900$ kg/m$^3$, dynamic viscosity $\mu_1 =$ $1024 \text{ mPa} \cdot \text{s}$, liquid–air surface tension $\gamma_1 = 50 \text{ mN/m}$, and liquid–water interfacial tension $\gamma_{13} = 39.4 \text{ mN/m}$. The interfacial tension between $L_1$ (laser oil) and $L_2$, $\gamma_{12} = 1.33 \text{ mN/m}$ with mineral oil and $\gamma_{12} = 6.75 \text{ mN/m}$ with Silicone Oil, is calculated using the interfacial tension formula[49] for nonpolar liquids as: $\gamma_{12} = \gamma_1 + \gamma_2 - 2\sqrt{\gamma_1 \gamma_2}$.

The core drop is chosen to have a greater density than the host liquid's density as otherwise, the encapsulated drop might not settle down to the bottom of the cuvette post-encapsulation. Also, the interfacial layer needs to be lighter than the host bath so that it can float on the free surface of the host bath. Further, the liquids were chosen such that the liquid-liquid encapsulation process remains thermodynamically feasible. In our previous work[37] we have shown that the condition for the encapsulated cargo to remain stable inside the host liquid is given by:

$$\gamma_{12} + \gamma_{23} - \gamma_{13} < 0 \qquad (1)$$

The combination of laser oil, mineral oil, and DI water, and laser oil, silicone oil, and DI water as L1, L2, and L3 satisfy the thermodynamic conditions given by equations (1).

Further, any liquid combination that are in direct contact (e.g., the core drop-interfacial liquid pair and the interfacial layer-host bath pair) at any time during the encapsulation process should be physicochemically compatible (i.e., mutually immiscible/non-reactive) with each other.

The experiments were conducted in a glass cuvette (Product Code: SC-02, Krüss GmbH, Hamburg, Germany) of inner dimension 36 mm x 36 mm x 30 mm with 2.5 mm wall thickness. Polished and passivated stainless-steel disposable needle tips with gauge 14 and inner diameter of 0.060" (Part No. 7018035, Nordson EFD, East Province, RI, USA) mounted on 1 ml NORM-JECT® sterile luer-slip syringes (Henke-Sass, Wolf GmbH, Tuttlingen, Germany) were used for dispensing the liquids.

In this method, the cuvette is filled with 20 ml of host liquid (DI water). The interfacial liquid is dispensed on the air-water interface using a pipette in a slow, controlled way to prevent any disruption of the interface, which forms a floating liquid lens on the water surface. The core drop (laser oil) is dispensed using the needle-syringe assembly mounted on a dispenser from an impact height H measured as the vertical separation between the needle tip and the surface of the interfacial layer. The generation of core drop involves a gradual increase in volume until it outweighs the liquid-air surface tension at the needle tip. The volume is dependent on the outer diameter of the needle, the liquid-air surface tension, and the effect of gravity[50] and thus remains invariant if the same class of needle tip and liquid combination is used. For validation of this claim dispensed volumes for multiple drops were initially measured with error bars. The average volume of core drop for laser oil was measured to be 15.5 µl with a standard deviation of 0.8 µl. Assuming a spherical geometry, the average volume corresponds to a drop radius of 1.54 mm which is below the system's capillary length scale. Assuming a very thin interfacial film wrapped around the core droplet, we can consider the apparent density of the core drop and the interfacial tension of the interfacial liquid with water to calculate the capillary length scale. Note that the aforementioned case of an encapsulated drop with an ultrathin wrapping shell layer where the density of the compound droplet could be approximated by the density of the core liquid corresponds to the worst-case scenario, with the lowest possible capillary length of the system. Any further increase in shell thickness corresponds to a consequent decrease in the effective density of the compound cargo and a resulting increase in capillary lengthscale. Therefore, if the size of the core drop is lower than this lowest achievable value of capillary length, it is logical to conclude that the capillary forces are dominant over the influence of gravity. The underwater capillary length scale, $l_{c,w}$ was given by:

$$l_{c,w} = \sqrt{\gamma_{23}/(\rho_1 - \rho_3)g} \qquad (2)$$

For mineral oil, the capillary length scale was 1.91 mm, while for silicone oil, it was 1.68 mm. Hence the core drop size was below the capillary length scale.

The interfacial liquid volume $V_{film}$ was varied from 150 µl to 750 µl with steps of 100 µl and the impact height $H$ of the droplet was varied from 6 cm to 15 cm. These experiments were repeated for the two interfacial liquids. The schematic representation of the liquid-liquid encapsulation process is shown in Figure 1(a). The interfacial dynamics of the encapsulation process were captured using a high-speed camera (Photron, FASTCAM Mini AX200) at 6400 fps connected to a Zoom Lens (Navitar 7000 Zoom with an effective focal length of 18–108 mm). The used high-speed camera has an internal memory of 32 GB. At 6400 fps, the maximum allowed recording window is 3.413 s at full resolution (1024 px × 1024 px). This is sufficient to record the complete time-resolved dynamics of encapsulation, usually finished within 500 ms. The captured data was transferred to a connected personal computer using the high-speed Gigabit ethernet connection between the camera and the computer for further analysis.

**Results and Discussion**

The process is characterized by the impact Weber number($We_i$) defined as $\frac{\rho_1 v^2 R_c}{\gamma_1} \approx \frac{\rho_1 2gHR_c}{\gamma_1}$, where $v$ is the velocity of the core drop just before the impact, $R_c$ is the radius of the core drop, $g$ is the acceleration due to gravity, $H$ is the impact height, and $\gamma_1$ is the interfacial tension of the core drop with air. Beyond a critical $We_i$,[38] the cargo detaches from the interfacial layer and settles at the bottom of the cuvette. The experiments were performed for an impact height of 6cm ($We_i = 69$), 9cm ($We_i = 103$), 12cm ($We_i = 138$) and 15cm ($We_i = 173$). The host liquid L3 is DI water and core drop liquid L1 is laser oil. The two interfacial liquids chosen were silicone oil and mineral oil and their dispensed volume was varied between 150 µl and 750 µl with steps of 100 µl. The radius of the core drop was kept constant at $R_c = 1.54$ mm.

The resulting interfacial dynamics involve the core drop dragging the interfacial liquid into the water pool, leading to the formation of a transient cylindrical interfacial liquid column with an air cavity inside, as can be seen in Figure 1(b). The expansion and subsequent contraction of the column can be seen in the temporal image sequence in Figure 1(c). The time $t = 0$ is defined as the instant when the maximum diameter of the core drop is at the interfacial layer free surface. Post-closure of the cavity a solid cylindrical column of interfacial liquid is formed which undergoes a slow thinning process.

The zoomed view of the timestamp at $t = 9.8$ ms in Figure 1(c) is shown in Figure 1(b). The critical forces acting on the liquid column and the important dimensions relevant to this study are demonstrated. The diameter of the liquid column during contraction (necking diameter), is represented as $D_N$, and its location from the $L_2$-$L_3$ (closure depth) is represented as $z_c$ shown with red arrows. The descent of the core drop is assisted by gravitational force $F_G$ while the deformed $L_2$-$L_3$ interface applies a restorative force $F_\gamma$, in the opposing direction. Additionally, the momentum of the core drop is also dissipated due to viscous forces $F_V$ exerted by the interfacial liquid. The competing forces, shown with green arrows, lead to the formation of the neck and final detachment, which will be discussed in detail in the following sections. The inner interface of the interfacial layer enclosing the cavity is also observed and marked with a yellow dotted line on one side for better viewing. The associated triple contact line with this interface can also be observed between the core drop, interfacial liquid, and air, shown with yellow arrows.

The interfacial dynamics can be resolved into 2 prominent stages:- Cavity formation and closure, and slow thinning of the interfacial liquid. For a better understanding of the process, two-time instants are defined:- $t_{\text{reach}}$ and $t_{\text{close}}$. The instant of cavity closure is recorded as $t_{\text{close}}$ (For example, $t_{\text{close}} = 17.5$ ms in Figure 1(c)). The closure depth, $z_c$ is defined as the distance between the neck and the free surface of the floating interfacial layer at the instance of cavity closure, i.e., at $t = t_{\text{close}}$. The instant $t_{\text{reach}}$ refers to the frame when the equator of the core drop (i.e., the location of the maximum width of the droplet along the horizontal axis) passes through the closure location. This is demonstrated using the dotted red horizontal line in Figure 1 (c) which marks the location of the neck at $t = t_{\text{close}} = 17.5$ ms. The vertical separation between the free top surface of the interfacial layer and the red dotted line, denoted by $z_c$, corresponds to the closure depth. The dotted red line is traced back to the previous frames to determine the instant when the equator of the core drop i.e. its maximum width coincides with the dotted line. This instant is defined as $t_{\text{reach}}$(for example, $t_{\text{reach}} = 7.3$ ms in Figure 1(c)). The instant $t_{\text{close}}$ also marks the end of the cavity dynamics phase and the beginning of the slow thinning phase. The cavity dynamics stage is further discussed in detail for varying experimental parameters. A qualitative discussion on the details of the thinning stage is provided in the supporting text S2 and S3.

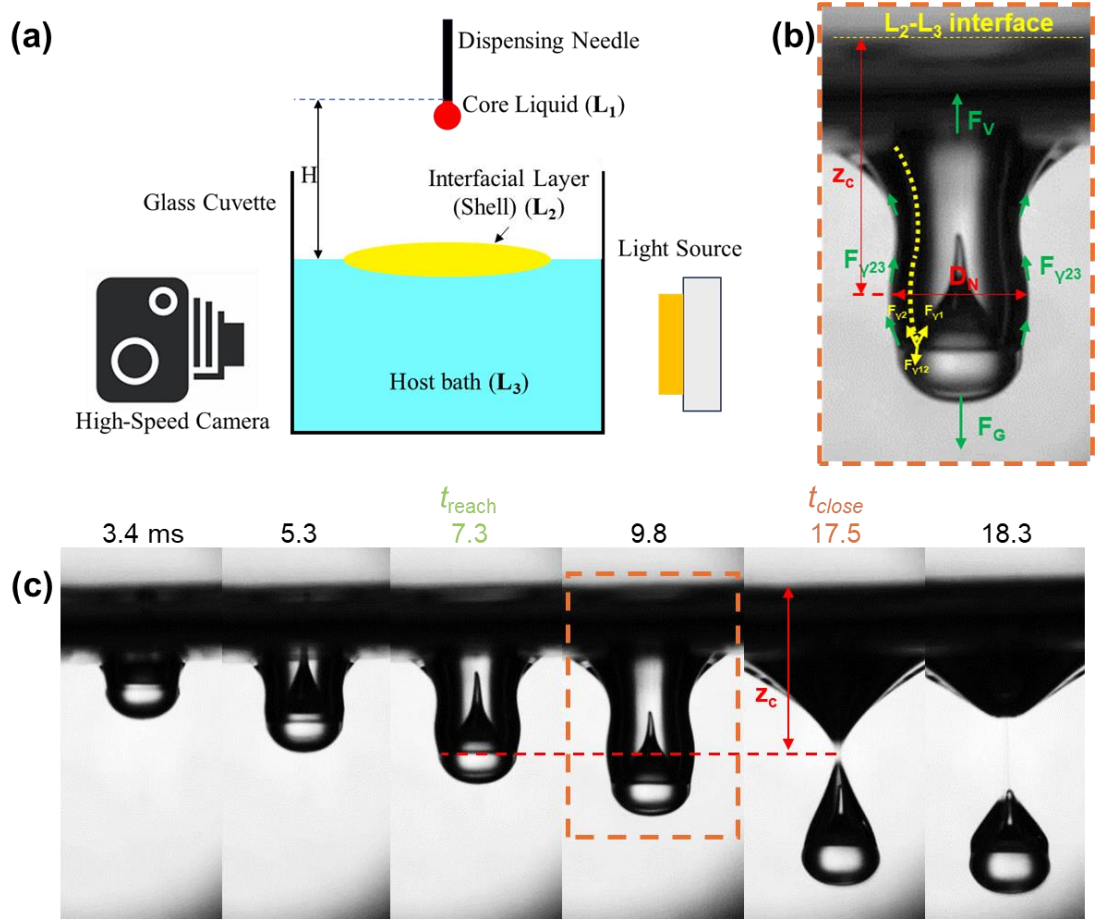

Figure 1. The schematic of the impacting droplet on a two-fluid system and the resulting dynamics are shown. (a) Core liquid drop (L1) is dispensed from an impact height of $H$ onto the floating interfacial layer (L2) which forms the shell of the encapsulated cargo. The interfacial layer is floating over the host liquid bath (L3). The interfacial dynamics is captured using a high-speed camera. (b) Schematic representation of the competing forces and the critical geometric parameters of the neck in terms of neck diameter ($D_N$) and closure depth ($z_c$) at $t = 9.8$ ms snap of Figure 1(c). (c) Interface evolution involved in the phenomenon for an interfacial layer of silicone oil ($V_{film} = 150$ μl) at $We_i = 138$ shown. The instant of cavity closure is denoted by $t_{close}$ is shown in orange. The red dotted line, drawn at a distance of $z_c$ from the free surface, is used to backtrack to the previous frames to determine the instant when the equator of the drop coincides with the line. This instant is denoted as $t_{reach}$ shown in green.

Figure 2 shows the comparative temporal evolution of the interface due to (a) the impact of a laser oil core droplet into the water pool in the absence of an interfacial layer, (b) the impact of a laser oil core droplet into the water pool in the presence of silicone oil ($V_{film} = 150$ μl, $We_i = 138$) interfacial layer, (c) the impact of a laser oil core droplet into the water pool in the presence of mineral oil ($V_{\text{film}} = 150\ \mu l$, $We_\text{i} = 138$) interfacial layer, and (d) the impact of a hydrophobic sphere into the water pool.[51] In the absence of an interfacial layer in Figure 2(a) the cavity formation was not observed but rather a crater formation was observed. While, in the presence of a thin film of the interfacial layer, the cavity formation and its subsequent collapse are observed as shown in Figures 2 (b) and 2(c). Therefore, introducing a two-fluid system leads to a significant change in interfacial dynamics.

In the case of silicone oil (see Figure 2(b)), the cavity was observed to slightly expand initially against the hydrostatic pressure from the water bath and the interfacial pull from the bulk interfacial layer to a maximum diameter followed by a rapid contraction and a neck formation. In the case of mineral oil (see Figure 2(c)), the expansion stage was not observed, and the neck diameter decreased with time. Thus, the cavity dynamics stage can be further resolved into the expansion and contraction stage. The time instants, $t_{\text{reach}}$ and $t_{\text{close}}$ are shown in green and orange respectively for both cases. The closure time ($t_{\text{close}} - t_{\text{reach}}$) is also lower for mineral oil (7.2 ms) compared to silicone oil(10.2 ms), indicating a higher interfacial pull for mineral oil. It can also be observed that the distance between the core drop location and the closure depth is higher in silicone oil compared to mineral oil leading to a sufficient amount of air being entrapped along with the core drop in the former case. Similar cavity formation and closure were reported by Quetzeri-Santiago et al. and Kroeze et al. in their study of accelerated jet impact on capillary bridges and deep pool respectively.[52,53] The cavity closure was no seal type at lower Weber numbers transitioning to deep seal type with increasing Weber numbers, and changed to surface seal type at higher Weber numbers. The phenomenon we observe in our study is mostly deep seal-type closure except for a few cases where no seal type is observed. From Figure 2(b) and 2(c) it can be implied that for the same volume of interfacial liquid and the same amount of kinetic energy possessed by the core drop, mineral oil provided more viscous resistance upon impact thereby reducing the core drop velocity despite mineral oil having a lower dynamic viscosity. Thus, the spread of floating interfacial liquid lens also plays a role in the cavity dynamics as well which is discussed further in the next subsection. For the same $V_{\text{film}}$, lower spreading would lead to higher film thickness leading to increased viscous dissipation.

The cavity dynamics is analogous to the reported literature on the impact of a hydrophobic solid sphere on the water surface as shown in Figure 2(d). However, the key difference is the presence of the interfacial layer which provides an additional capillary pressure along with the hydrostatic pressure to the cavity. Therefore, the expansion stage which is always prominent for the impact of a solid sphere may not exist in the current system. Post necking closure, the column underwent a thinning process(see right-most image sequence in Figure 2(b) and 2(c)), and eventually, detachment occurred. Hence, the second key difference is the presence of a thinning stage of the interfacial liquid post cavity closure.

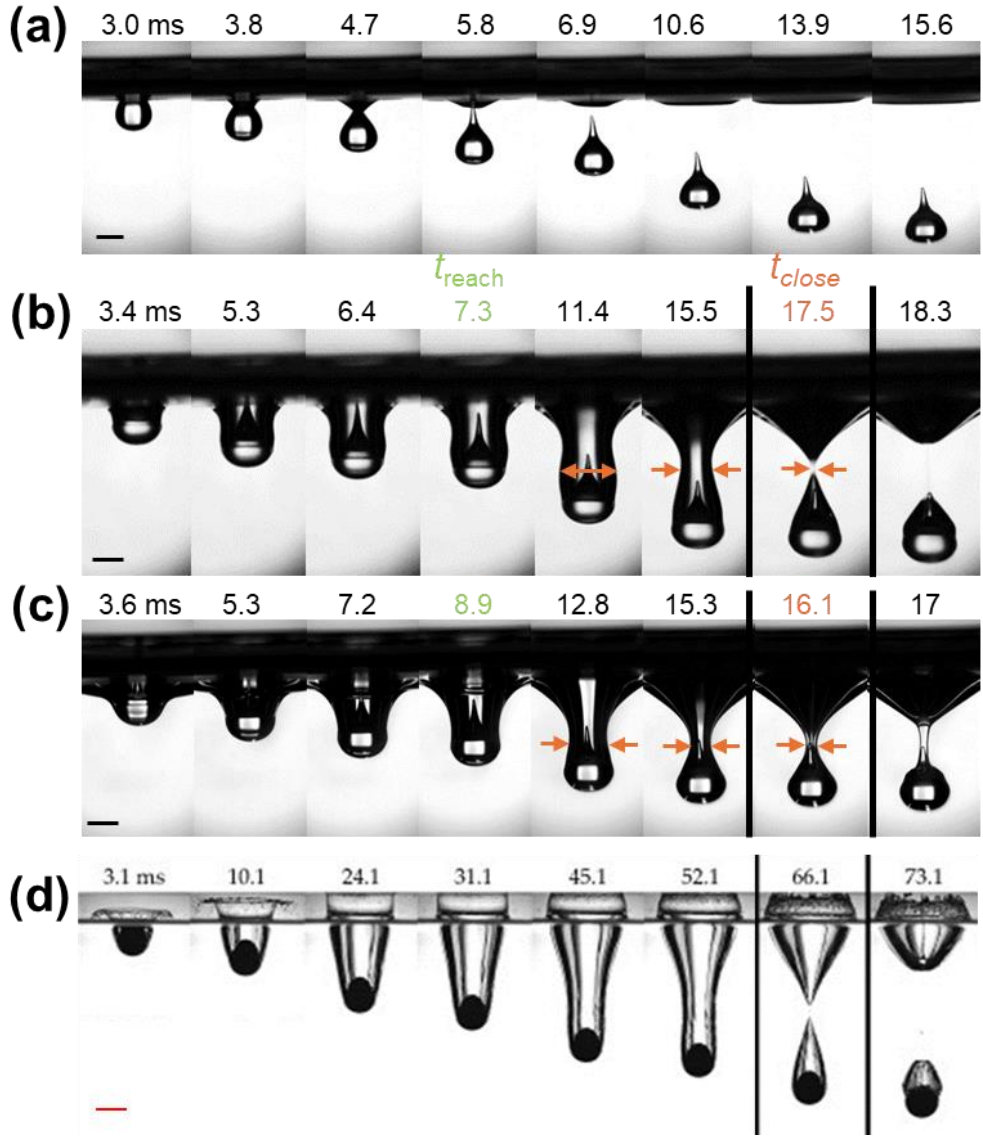

Figure 2. The interface evolution due to the impact of laser oil on the water pool in the presence and absence of the interfacial liquid is shown. The cavity dynamics due to the impact of a hydrophobic solid sphere into the water pool as reported in the literature is shown for comparison. (a) Interface evolution due to the impact of a laser oil drop in the absence of an interfacial layer ($We_i = 138$). Interface evolution due to the impact of a laser oil drop in the presence of silicone oil (b) and mineral oil (c) interfacial layer ($V_{film} = 150$ μl, $We_i = 138$). Scale bar (black) 2mm. (d) Interface evolution due to the impact of a Teflon sphere into the water pool with an impact speed of 2.17 ms⁻¹. Scale bar(red) 2.54 cm. Reproduced with permission.[51] 2024, AIP Publishing.

Expansion and Contraction stages

A detailed study of the expansion and contraction stages is important because the governing necking dynamics determine if any air will be entrapped along with the core drop. The core drop, upon impact, loses a part of its kinetic energy due to viscous dissipation by the interfacial layer. The dissipation is higher for higher thickness which increases with the volumes of interfacial liquid. Additionally, it also depends on the lateral spread of the interfacial layer on the host bath, as lower spreading leads to the formation of a thicker floating liquid lens for the same volume of the interfacial liquid, which causes more energy dissipation compared to a thinner floating liquid lens. The spreading parameter of the interfacial layer on the water was calculated as $S_{23} = \gamma_3 - \gamma_2 - \gamma_{23}$ for silicone oil is 16.3 mN/m and for mineral oil is 12 mN/m, thus the spreading is higher for silicone oil on water compared to that of mineral oil. Also due to the positive spreading parameter, the oil film forms a pseudo-partial wetting state.[54] A very thin film of oil spreads, covering the entire water surface, and the remaining volume forms a lens. Thus, measuring the lens diameter or thickness is non-feasible.

The thickness of the interfacial layer is lower for silicone oil compared to mineral oil for the same $V_{\text{film}}$. Hence, the initial viscous dissipation upon impact is also lower in silicone oil compared to mineral oil despite the former having higher dynamic viscosity. Due to the higher loss of kinetic energy in mineral oil, the core drop from an impact height of 6 cm ($We_{\text{i}} = 69$) could not penetrate the interfacial layer and remained trapped in the interfacial layer even at the lowest explored $V_{\text{film}} = 150$ μl. In contrast, in silicone oil, interfacial penetration and subsequent encapsulation were observed up to $V_{\text{film}} = 550$ μl of interfacial liquid and the core drop was trapped at $V_{\text{film}} = 650$ μl of interfacial liquid. Further, due to higher spread and lower viscous dissipation in silicone oil, despite interfacial penetration, air entrapment was seen to occur for an impact height of 15 cm ($We_i = 173$) even with the highest studied $V_{film} = 750$ µl of interfacial liquid, while in the case of mineral oil, air entrapment was observed below $V_{film} = 450$ µl. Thus, for a comparative study to understand the exact dynamics, we have chosen the intermediate impact heights of 9 cm and 12 cm for further analysis. The transient variation of the non-dimensional neck diameter $D^* = D_N/2R_C$ of the interfacial liquid column at the closure depth, $z_c$ is plotted for various cases and shown in Figure 3 (a), (b) and (c). Since cavity closure dynamics have been reported to be a radial phenomenon,[51,55,56] thus the interfacial column diameter was tracked at a constant vertical depth, specifically the closure depth. The cavity dynamics has been reported to follow a Rayleigh-type equation.[53,56,57] A simple derivation and scaling of the governing Euler equation is given in the supporting text S1. The inertia-capillary time scale given by $\tau_c = \sqrt{\rho_3 R_c^3/(\gamma_2 + \gamma_{23})}$ was calculated to be 7.8 ms for the case of silicone oil and 8.1 ms for the case of mineral oil which is in the range of our experimental timescale. The temporal axis was non-dimensionalized as $\tau^* = (t - t_{reach})/\tau_c$. Hence, $D^*$ was plotted from $\tau^* = 0$ to $\tau^* = (t_{close} - t_{reach})/\tau_c$ for $We_i = 103$ and $We_i = 138$ for different $V_{film}$ of mineral oil and silicone oil. The dynamics was observed to be different in different cases due to variations in the competing forces in each case. A case-by-case discussion is given in the following sections:

Silicone Oil as an interfacial layer:

In Figure 3(a), which shows the neck diameter evolution for $We_i = 138$ for silicone oil as an interfacial layer, it can be seen that the neck diameter initially expands slightly and reaches a maximum diameter and then contracts due to interfacial pull and the hydrostatic pressure. This duration and extent of the expansion stage reduces with increasing $V_{film}$ and beyond $V_{film} = 450$ µl, the expansion stage does not exist. The interfacial liquid at lower $V_{film}$ has little viscous as well as interfacial resistance to offer to the core drop and an inertia-driven rapid contraction of the neck is observed. Similar necking diameter evolution has been reported in the literature where a constant velocity disk was dragged down a water bath forming a void of air column.[56] The reported expression in non-dimensional form solving the Rayleigh-equation is given by:

$$D^* = \sqrt{{D_{\max}^*}^2 - \beta(\tau^* - \tau_{\max}^*)^2} \qquad (3)$$

where $\tau^* = (t - t_{\text{reach}})/\tau_c$ is the measured time adjusted for $t_{\text{reach}}$ and non-dimensionalized with inertia-capillary timescale $\tau_c$, $D_{\max}^*$ is the maximum, $\tau_{\max}^*$ is the time where $D^* = D_{\max}^*$, i.e. the peak diameter achieved during the expansion stage, and $\beta$ is a free parameter of the model, which depends on the Bond number and the closure depth. A derivation is given in the supporting text S1. Similarly, for the rapid contraction stage, the reported expression is

$$D^* = C\sqrt{\tau^* - \tau_{\text{close}}^*} \qquad (4)$$

where $\tau_{\text{close}}^* = (t_{\text{close}} - t_{\text{reach}})/\tau_c$ is the non-dimensionalized time of neck closure adjusted for $t_{\text{reach}}$ and $C$ is a free parameter of the model. Using the above expressions, best-fitted curves were plotted (shown in solid line) against the experimentally obtained data (shown with dots) in Figure 3(a). For $V_{\text{film}}$ of 550 µl and 650 µl, the experimental plot was observed to be flat in the beginning followed by a contraction. In this case, only the contraction section was plotted, based on Eq. (2). In all the cases the reported expression was found to be highly correlated with experimental data. The free parameter $\beta$ was calculated to be 3.477 and $C$ varied between 1.265 to 1.327.

The expansion of the column occurs because when the core drop traverses down it displaces the water underneath itself to the side. It can be argued that the radial velocity of the water displaced is proportional to the core drop velocity.[56,57] Thus, the radial expansion is also higher with higher core drop velocity. Since increasing $V_{\text{film}}$ leads to higher viscous dissipation, the core drop velocity decreases and thus the maximum neck diameter during the expansion stage also decreases. This initial radial expansion occurs against the hydrostatic pressure from the water pool as well as the interfacial pull from the interfacial liquid layer. When the core drop velocity is low due to higher viscous dissipation at higher $V_{\text{film}}$ the radial expansion is limited to a small depth of the water pool above the closure depth and by the time the core drop reaches the closure depth ($\tau^* = 0$), the inertial radial expansion has already been countered by the hydrostatic pressure and interfacial forces. Therefore, at $V_{\text{film}} = 550$ μl, the initial stage of the $D^*$ vs. $\tau$ plot (see Figure 3(a)) is observed to be almost flat, and at $V_{\text{film}} = 650$ μl the neck diameter is observed to follow a slight decrease before the rapid collapse. The inset shows the zoomed plot for 150 μl, 550 μl, and 650 μl for $\tau^* = 0$ ms to $\tau^* = 0.3$, where the liquid column can be seen expanding for 150 μl while directly contracting for 650 μl. The higher core drop velocity also leads to a higher vertical difference between the location of closure depth and the core drop at the time of neck closure which corresponds to a higher volume of air entrapped. Thus, the volume of air entrapped decreases with increasing $V_{\text{film}}$ until the core drop location is near the closure depth at the time of neck closure, beyond which no air is entrapped, which was observed for $V_{\text{film}} = 650$ μl.

Similarly, the neck evolution at the lower $We_{\text{i}}$ (103) can be explained as shown in Figure 3(b), where the initial expansion of the neck is non-existent, and the non-dimensional neck diameter seems to be invariant with $V_{\text{film}}$ as almost overlapping curves are obtained. The initial low velocity of the core drop fails to cause any expansion as the interfacial pull dominates over the inertial expansion. The curve fitting with the reported expression also fails in this case. The time of cavity closure is observed to be around 0.73 while the fastest closure time with $We_{\text{i}} = 138$ was at 0.965 ($V_{\text{film}} = 650$ μl), which in dimensional terms is 1.8 ms faster. At lower velocities, air entrapment was not observed beyond 350 μl.

<u>Mineral Oil</u> as an interfacial layer:

Mineral oil has a lower spread, forms a thicker floating lens, and offers a higher viscous dissipation. Thus, it was observed that even with a higher $We_i$ (138), the expansion stage is non-existent as shown in Figure 3(c). It is observed that for $V_{film}$ of 150 µl to 450 µl, the collapse was observed to be faster with increasing $V_{film}$. However, the duration of cavity closure seemed to be coinciding for 450 µl, 550 µl and 650 µl. At 750 µl, the graph is observed to be flatter with $D^* = 0.4$ at cavity closure, which is significantly greater compared to the other cases, where $D^*$ was around 0.2 to 0.25, indicating a thicker interfacial column formation for the thinning stage. When the velocity of the core drop is greatly reduced the capillary forces start dominating and the cavity closure dynamics change from deep seal to no seal[52,53,57] as shown in the temporal image sequence in Figure 3 (d) ($We_i = 103$). The triple contact line progresses along the core drop surface and completely covers the core drop as the air cavity is pushed out. The end of the contraction stage can be assumed to be instant when the core drop is completely wrapped with the interfacial liquid and its duration is observed to be small (0.59), beyond which a slower thinning of the interfacial column follows. Thus, for the same impact Weber number of 103, deep seal closure was observed for the case of silicone oil while no seal was observed for the case of mineral oil implying that the spread of the interfacial liquid can dictate the type of cavity closure dynamics.

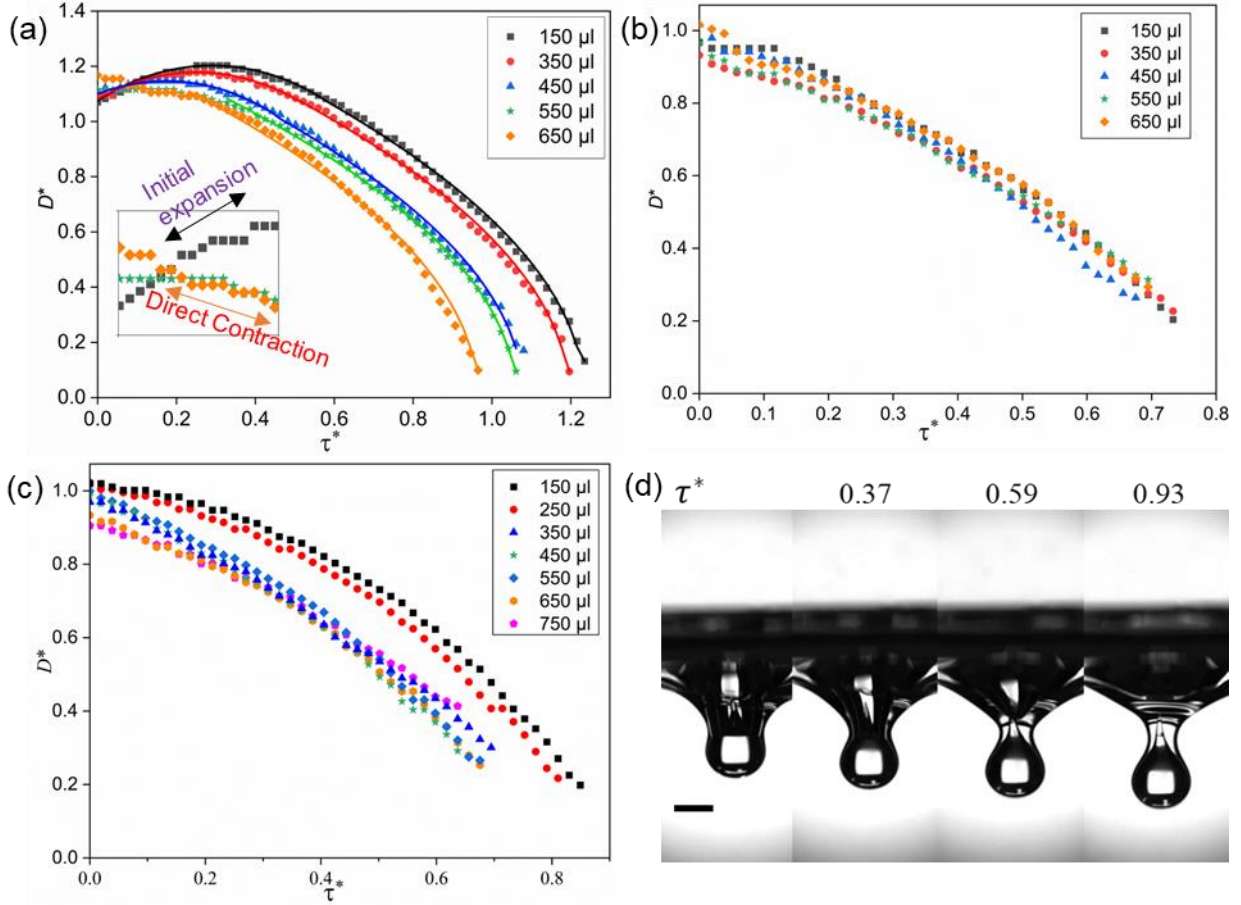

Figure 3. The temporal variation of non-dimensional neck diameter $D^* = D_N/2R_C$ (during the contraction stage is shown for silicone oil as an interfacial layer for $We_i = 138$ (a) and $We_i = 103$ (b), and for mineral oil as an interfacial layer for $We_i = 138$ (c) for different $V_{film}$. The inset in (a) shows the zoomed plot of $D^*$ for 150 µl, 550 µl and 650 µl from $\tau^* = (t - t_{reach})/\tau_c = 0$ to $\tau^* = 0.3$. The interface evolution in the contraction stage for $We_i = 103$ for mineral oil($V_{film} = 350$ µl) is shown in Figure 3(d). Scale bar 2 mm.

The image sequence for silicone oil for $We_i = 138$ with (a) $V_{film} = 150$ µl (b) $V_{film} = 550$ µl, and (c) $V_{film} = 650$ µl is shown in Figure 4. It can be seen that in Figure 4(a), the cavity was expanding before $\tau^* = 0$ (i.e., $t = t_{reach}$) and kept expanding after that. In the second case (see Figure 4(b)), except for the initial dome shape at $t = 3.3$ ms, the column diameter was almost constant, thus implying that the inertial radial push was at equilibrium with the interfacial forces and the hydrostatic pressure. In the third case (see Figure 4(c)), the interfacial forces dominated and didn't allow for any radial expansion. This is a major difference compared to the impact of a solid sphere, wherein an expansion of the column is visible[51]. The presence of the interfacial layer introduces an extra capillary pressure in addition to the hydrostatic pressure that prevents the inertial radial push for lower core drop velocity.

The interfacial dynamics comparing these three cases have been given in supporting video S1. Further for mineral oil, for $V_{film} = 350$ µl, the interface evolution for $We_i = 103$ and $We_i = 138$ is shown in supporting video S2. It could be seen that the cavity formation is of no seal type and contraction is of shorter duration followed by a slower thinning stage for $We_i = 103$. From these observations, it can be concluded that a deep seal type cavity closure dynamics with purely inertial closure was observed at higher impinging drop velocity which transitioned to an inertia-capillary closure with decreasing velocity. With further decreasing velocity, the deep seal type closure transitioned into a no seal type closure(see Figure 3(d))

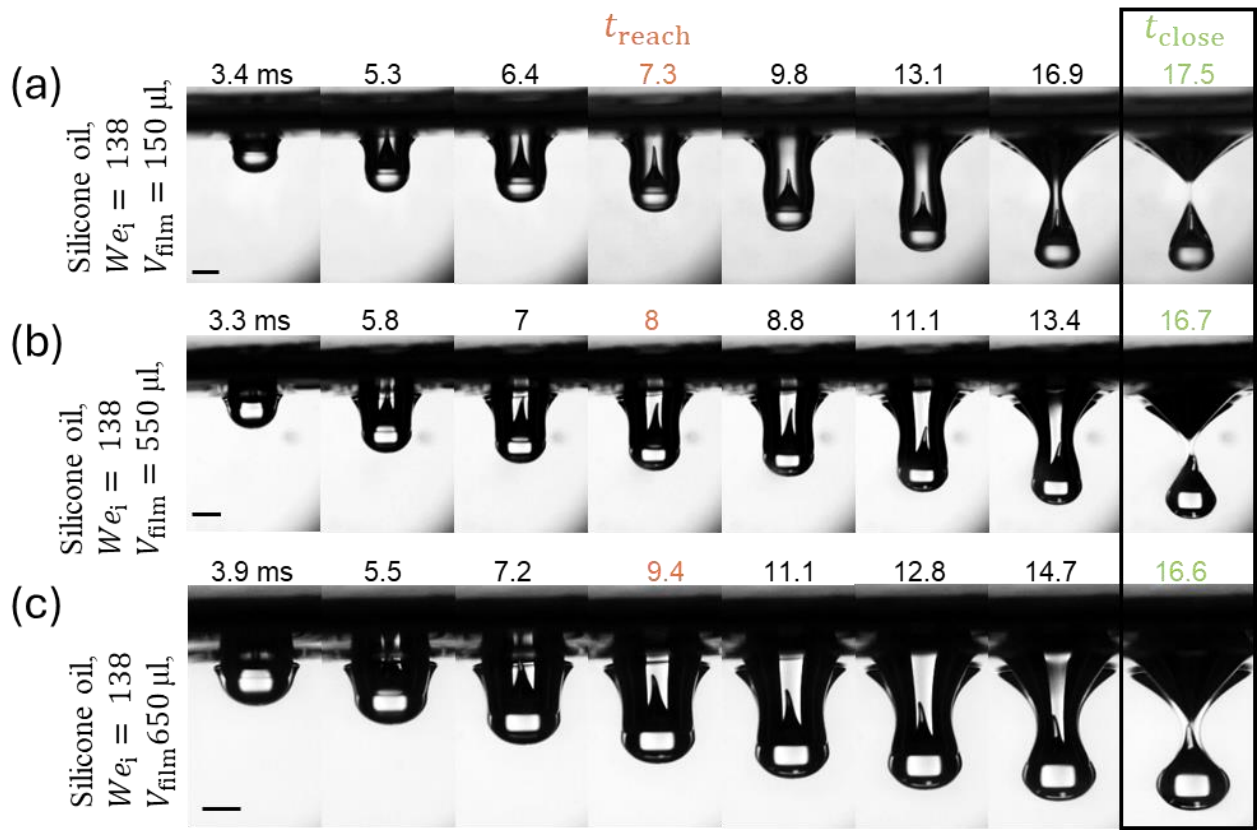

Figure 4. The image sequence for silicone oil as an interfacial layer for $We_i = 138$ for three different volumes are shown. (a) $V_{film}$ =150 µl,(b) $V_{film}$ =550 µl, and (c) $V_{film}$ =650 µl. The interface evolution shows an expansion and contraction in Figure 4(a), which remains constant followed by a contraction in Figure 4(b) and only a contraction in Figure 4(c). Scale bar 2 mm.

**Conclusion**

In conclusion, we investigated the necking dynamics and the eventual detachment of cargo in a four-fluid interface (core liquid, interfacial liquid, host liquid, and trapped air phase) in a liquid-liquid encapsulation process. The dynamics can be divided into a cavity expansion and contraction stage followed by slow thinning. The core drop, upon impact, dragged the interfacial liquid along with it into the water pool forming an interfacial fluid cavity with air in the middle. This cylindrical column under the influence of interfacial pull from the bulk of the liquid and hydrostatic pressure from the water pool, underwent a rapid collapse which we are classifying as the contraction stage. Following the collapse, a cylindrical fluid column remains which undergoes a thinning stage and eventually pinches off from the encapsulated cargo. The contraction stage was observed to be dependent on the spread of the interfacial liquid at the water interface and the impact height of the core drop. Silicone oil owing to its higher spreading compared to mineral oil formed a thin floating lens and thereby offered lesser viscous dissipation upon impact. The viscous dissipation also increases with an increase in interfacial volume. The necking dynamics were observed to be purely inertial at low interfacial volumes of silicone oil due to lower viscous dissipation. The neck was observed to have an initial expansion as well before the collapse. With the increasing interfacial volume of silicone oil, using lower spread mineral oil with lower volumes or decreasing impact height, the velocity of the core drop could be decreased wherein the interfacial pull from the fluid column prevented the initial expansion of the neck and also caused faster collapse thus an inertia-capillary necking dynamics was observed. The cavity closure is observed to be deep seal type. Upon further increasing the viscous dissipation by using higher interfacial volumes of mineral oil, the contraction stage was observed to be negligible as the air column was driven out simultaneously, as the core drop progressed i.e. a no seal type closure was observed. The resulting interfacial fluid column underwent a slow thinning process.

**Supporting Information**

Supporting Text S1 Cavity Dynamics

Supporting Text S2 Thinning Stage

Supporting Text S3 Non-Dimensional neck diameter versus time for entire cavity closure and thinning stage

Supporting Video S1: Effect of interfacial layer volume on the interface dynamics with silicone oil as the interfacial liquid.

Supporting Video S2: Effect of kinetic energy of Impact on the interface dynamics with mineral oil as the interfacial liquid.

**Conflicts of Interest:**

The key Liquid-Liquid Encapsulation technology led to the formation of a startup with a presence in Canada (Scalable Liquid Encapsulation Inc.) and the Netherlands (SLE Enterprises B.V.). Both S.M and S.K.M have equity stakes in the startup.

**Acknowledgements:**

A.C performed the experiments and analyzed the data with help from S.M. S.M developed the image processing framework for the quantitative analysis of interface dynamics from the high-speed experimental videos. A.C wrote the paper with inputs from S.M. S.M, and S.K.M conceived the study. S.K.M supervised the project with support from S.M. All authors contributed to the revision and subsequent improvements of the manuscript. The work was supported by the S.K.M's Discovery Grant (RGPIN-2024-03729) from the Natural Sciences and Engineering Research Council (NSERC), Canada, and the startup grant from the University of Waterloo. A.C.

acknowledges the financial support from the Waterloo Institute for Nanotechnology (WIN), University of Waterloo in the form of Nanofellowship 2023.

# Interface Dynamics at a Four-fluid Interface during Droplet Impact on a Two-Fluid System

*Akash Chowdhury, Sirshendu Misra, Sushanta K. Mitra*[*]

Micro & Nano-Scale Transport Laboratory, Waterloo Institute for Nanotechnology, Department of Mechanical and Mechatronics Engineering, University of Waterloo, 200 University Avenue West, Waterloo, Ontario N2L 3G1, Canada

Email: skmitra@uwaterloo.ca

This file includes:

- Supporting Text
    1. Cavity Dynamics
    2. Thinning stage
    3. Non-Dimensional neck diameter versus time for entire cavity closure and thinning stage
- Figures
    1. Figure S1
    2. Figure S2
- Videos
    1. Video S1: Effect of interfacial layer volume on the interface dynamics with silicone oil as the interfacial liquid.
    2. Video S2: Effect of kinetic energy of Impact on the interface dynamics with mineral oil as the interfacial liquid.

**Supporting Text**

**S1. Cavity Dynamics:**

The cavity dynamics have been reported to be a purely radial motion.[1,2] We consider the domain of host liquid with its boundary covered by a thin interfacial liquid film that is exposed to the air cavity. The radial component of the Euler equation in the cylindrical co-ordinate is given by:

$$\frac{\partial v_r}{\partial \tau} + v_r \frac{\partial v_r}{\partial r} = -\frac{1}{\rho_3} \frac{\partial p}{\partial r} \tag{1}$$

where $v_r$ is the radial velocity component in the host liquid, $\tau = t - t_{reach}$ is the time adjusted for $t_{reach}$, $p$ is pressure and $\rho_3$ is host liquid density. The continuity equation and boundary conditions on the water-film interface is given by:

$$r v_r = R_N \dot{R_N} \tag{2}$$

where $R_N = D_N/2$ is the radius of the cavity and $\dot{R_N}$ is the velocity of the cavity wall. We have assumed that there is no loss of velocity in the thin interfacial film and the film thickness remains constant. Substituting (2) in (1) we get:

$$\frac{\partial}{\partial \tau}\left(\frac{R_N \dot{R_N}}{r}\right) + \frac{\partial}{\partial r}\left(\frac{v_r^2}{2} + \frac{p}{\rho_3}\right) = 0 \tag{3}$$

We integrate (3) over r from the cavity wall $R_n$ to a quiescent point in the host liquid. The pressure boundary conditions at the host liquid interface can be assumed to be the sum of capillary pressure from the host liquid-interfacial film interface and the interfacial film-air cavity interface given by $(\gamma_2 + \gamma_{23})\nabla.\hat{n}$, where $\gamma_2$ is the surface tension of the interfacial liquid and $\gamma_{23}$ is the interfacial tension of the interfacial liquid with the host liquid. We have assumed the curvature change over the thin film to be negligible. Further assuming large slopes of the cavity wall, such that the longitudinal component of the curvature is negligible, we get $\nabla \cdot \hat{n} = 1/R_N$ . We achieve a Rayleigh-type equation:

$$\frac{d(R_N \dot{R}_N)}{d\tau} \log\left(\frac{R_N}{R_\infty}\right) + \frac{1}{2}\dot{R}_N{}^2 = gz_c + \frac{(\gamma_2 + \gamma_{23})}{\rho_3 R_N} \tag{4}$$

The term $\log\left(\frac{R_N}{R_\infty}\right)$ has been reported to be in the $O(1)$. A scaling of the inertial terms in LHS with the surface tension term gives the inertia-capillary time scale as:

$$\frac{R_c^2}{\tau_c^2} \sim \frac{(\gamma_2 + \gamma_{23})}{\rho_3 R_c} \Rightarrow \tau_c = \sqrt{\frac{\rho_3 R_c^3}{\gamma_2 + \gamma_{23}}} \tag{5}$$

The time scale calculated for silicone oil is 8.1 ms while that for mineral oil is 7.8 ms which is in the range of our experimental time scale.

For the expansion stage of the purely inertial flow cases, the surface tension component was ignored. Additionally, ignoring $\dot{R}_N{}^2$ and assuming $\log\left(\frac{R_N}{R_\infty}\right) \approx \log\left(\frac{R_{max}}{R_\infty}\right)$ [1] we get

$$R_N^2 = R_{max}^2 + \frac{gz_c}{\log\left(\frac{R_{max}}{R_\infty}\right)}(\tau - \tau_{max})^2 \tag{6}$$

Rewriting (6) in terms of the cavity diameter and non-dimensionalizing with $D^* = R_N/R_c$, $D_{max}^* = R_{max}/R_c$, $\tau^* = \tau/\tau_c$ and $\tau_{max}^* = \tau_{max}/\tau_c$, we get:

$$D^* = \sqrt{D_{max}^*{}^2 - \beta(\tau^* - \tau_{max}^*)^2} \tag{7}$$

where $\beta = -\frac{\rho_3 g R_c z_c}{(\gamma_2 + \gamma_{23}) \log\left(\frac{R_{max}}{R_\infty}\right)} = -\frac{z^* Bo}{\log\left(\frac{R_{max}}{R_\infty}\right)}$ . The bond number, $Bo$ is defined as $\frac{\rho_3 g R_c^2}{(\gamma_2 + \gamma_{23})}$ and the non-dimensional closure depth, $z^*$ is defined as $\frac{z_c}{R_c}$.

The fast inertial contraction stage was given by[1]:

$$\frac{d(R_N \dot{R}_N)}{d\tau} = 0 \tag{8}$$

which gives:

$$D^* = C\sqrt{\tau^* - \tau_{\text{close}}^*} \tag{9}$$

Since the expansion stage is absent at lower Weber numbers and with interfacial higher volumes, and curve fitting with (9) and (8) fails for Figures 3(b) and 3(c) in the main text, the assumption of ignoring the capillary pressure term is not valid. It accelerates the process of necking and thus we can observe a shift of the cavity closure domain from purely inertial to inertia-capillary.

## S2. Thinning stage

Post-cavity closure ($t > t_{\text{close}}$), the thinning stage is observed which involves the stretching of a thin cylindrical column of interfacial liquid that is being subjected to interfacial pull from the bulk interfacial layer floating on the water surface from the upper end and the interfacial liquid wrapped around the core drop from the lower end. Depending on the impact Weber number, the drop morphology can be different leading to different phenomena. Owing to the highly viscoelastic nature of laser oil, the core drop possesses a long tail when it gets detached from the dispensing nozzle. During its downward traversal (in the air) towards the floating interfacial layer, the tail gets shortened, and the drop attains a pendant shape due to the effect of surface tension. Given a sufficient amount of time, the drop would attain a spherical shape due to surface tension. However, in the current setup, the core drop impacts the interfacial layer before it attains a spherical shape. The time gap, $t_{gap}$ between the detachment of the drop from the dispensing needle and its contact with the floating interfacial layer dictates the length of the tail before impact. Evidently, assuming gravity assisted free-fall of the core drop in air, $t_{gap} \sim H^{\frac{1}{2}} \sim We_i^{\frac{1}{2}}$. Therefore, the thin viscoelastic tail is expected to be prominent in the impinging core drop at a lower Weber number. The temporal interface evolution is shown in Figure S1 for different Weber numbers where $\tau' = t - t_{\text{close}}$. When it is impacted from a height of 6 cm ($We_i = 69$), it doesn't get the required time to form a pendant shape and is observed to have a long tail attached to it. As shown in Figure S1(a), the tail of the drop along with the wrapped interfacial layer is stretched until the encapsulated cargo has settled at the bottom of the cuvette. The liquid column is observed to undergo a slow thinning process until it forms a thin thread which ultimately detaches at multiple points. Since laser oil is viscoelastic, this thinning process is dominated by laser oil and the total duration of the stage was around 100ms. However, it is difficult to comment on whether the laser oil column was wrapped with the interfacial layer throughout the process or if the interfacial liquid had broken up before.

In the case of an impact height of 9 cm ($We_i = 103$), the size of the tail of the core drop, although shorter than in the previous case, varied between experiments. In case the tail size was slightly long it was observed to undergo stretching along with the interfacial layer as shown in Figure S1(b). However, unlike the previous case, slow thinning was not observed and consequently the duration of the stage was around 32 ms. Whereas if the core drop has a smaller tail, the tail doesn't influence the thinning process and only the interfacial liquid column above the tail portion is stretched and eventually detaches as shown in Figure S1(c). Since laser oil was not involved in this process, the time taken was 10.6 ms.

In the case of a higher impact height such as 12cm ($We_i = 138$) and 15cm ($We_i = 172$), the core drop is developed into a pendant shape and the location of closure depth is above the core drop hence the laser oil does not influence the thinning stage as shown in Figure S1(d). The interfacial liquid column undergoes stretching and eventually detaches at multiple points to form a satellite droplet. The duration of the process is also observed to be faster taking 6.8 ms.

The duration of the stage was observed to increase with $V_{film}$ implying the fact that increasing volume increases viscous dissipation and due to decreased core drop velocity, the thinning of the interfacial layer is also slower.

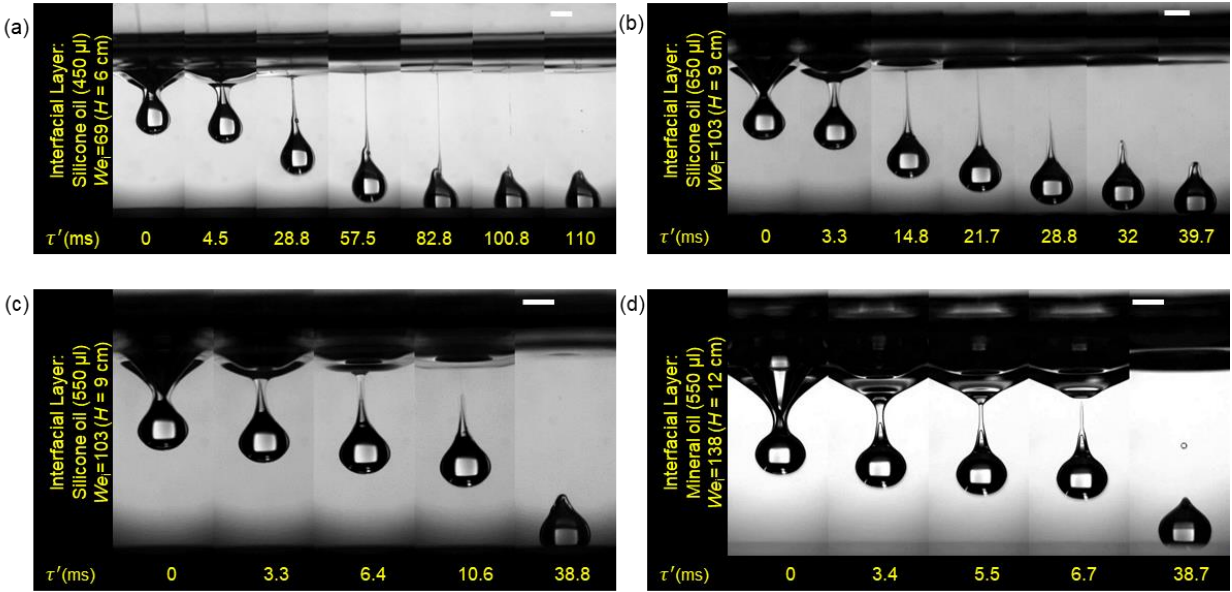

**Figure S1:** The time-series evolution of the thinning stage for various cases is shown. (a) $We_i =$ 69($H = 6$ cm and $V_{film} = 450$ μl of silicone oil as interfacial layer) (b) $We_i = 103$ ($H = 9$ cm and $V_{film} = 650$ μl of silicone oil as interfacial layer) (c) $We_i = 103$ ($H = 9$ cm and $V_{film} = 550$ μl of silicone oil as interfacial layer) (d) $We_i = 138$ ($H = 12$ cm and $V_{film} = 550$ μl of mineral oil as interfacial layer). Core drop can be observed to cause an influence in this stage, in case the drop is not developed into a pendant shape and its tail undergoes a thinning(a, b), while at a higher Weber number, only the interfacial layer is under play.

## S3: Non-Dimensional neck diameter versus time for entire cavity closure and thinning stage

The non-dimensional neck diameter $D^*$ plotted against the non-dimensional time $\tau^*$ for the entire range of contraction and thinning stages in the case of mineral oil as the interfacial liquid($V_{film} = 350$ μl) at $We_i = 103$ i.e. the case shown in Figure 3(d) in the main text is shown in Figure S2. The contraction stage and the thinning stage are demarcated by the $\tau^* = 0.59$ line which we have defined as the instant when the core drop is completely covered with the interfacial liquid. It can be seen that the curve undergoes a point of inflection at the end of the contraction stage. The neck diameter, while decreasing rapidly in the contraction stage, was observed to undergo a slow thinning stage with the duration being roughly 3.8 times that of the contraction stage.

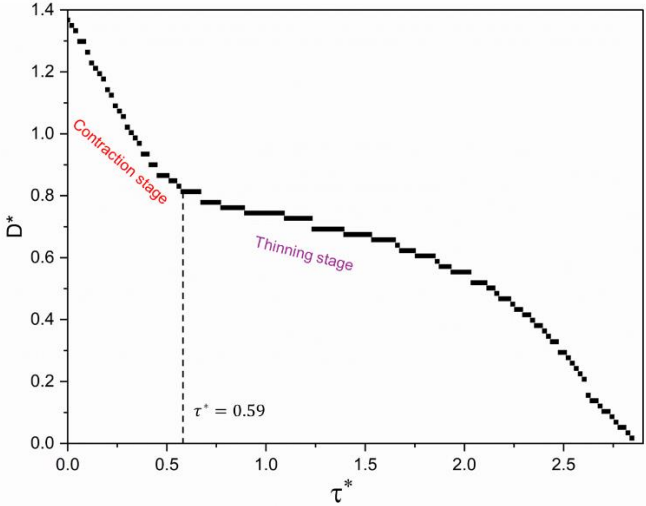

**Figure S2:** The temporal variation of non-dimensional neck diameter $D^* = D_N/2R_C$ for the entire duration of the contraction stage and the thinning stage is shown for mineral oil as an interfacial layer for $We_i = 103$ and $V_{film} = 350$ μl. The line $\tau^* = 0.59$ is denotes the end of the contraction stage where a point of inflection on the curve can be observed as well.

**List of Supporting Videos**

**Supporting Video S1:** Effect of interfacial layer volume on the interface dynamics with silicone oil as the interfacial liquid.

**Supporting Video S2:** Effect of kinetic energy of Impact on the interface dynamics with mineral oil as the interfacial liquid.